\documentclass[10pt,conference]{IEEEtran}
\IEEEoverridecommandlockouts
\usepackage{amsmath,amssymb,amsfonts}
\usepackage{graphicx}
\usepackage{textcomp}
\usepackage{xcolor}
\usepackage{booktabs}
\usepackage{algorithm}
\usepackage{pifont}
\usepackage{algpseudocode}
\usepackage{xspace}
\usepackage{subcaption} % 用于子图支持
\usepackage{multirow}
\usepackage{multicol}
\usepackage[breakable]{tcolorbox}
\usepackage{enumitem}
\usepackage{amsmath}
\usepackage{hyperref}
\usepackage{bbm}

\newcommand{\ie}{{\emph{i.e.}}\xspace}
\newcommand{\etc}{etc.}
\newcommand{\eg}{{\emph{e.g.}}\xspace}

\usepackage[numbers,sort&compress]{natbib}
\usepackage[normalem]{ulem}
\useunder{\uline}{\ul}{}
\usepackage{threeparttable}
\usepackage{makecell}
\usepackage{tcolorbox}

\usepackage{listings}

\usepackage{tikz}

\def\BibTeX{{\rm B\kern-.05em{\sc i\kern-.025em b}\kern-.08em
    T\kern-.1667em\lower.7ex\hbox{E}\kern-.125emX}}

\setlist[itemize]{leftmargin=10pt}

\newcommand*{\circled}[1]{\lower.7ex\hbox{\tikz\draw (0pt, 0pt)    circle (.3em) node {\makebox[1em][c]{\tiny #1}};}}

\renewcommand{\paragraph}[1]{\vskip 0.01in \noindent {\bf #1.}}

\newcommand{\greysum}[1]{}

\makeatletter
\DeclareRobustCommand\onedot{\futurelet\@let@token\@onedot}
\def\@onedot{\ifx\@let@token.\else.\null\fi\xspace}
 
\def\eg{\emph{e.g}\onedot} \def\Eg{\emph{E.g}\onedot}
\def\ie{\emph{i.e}\onedot} 
 
\def\etc{\emph{etc}\onedot}

\DeclareRobustCommand{\method}{CTM\xspace}
\DeclareRobustCommand{\fullmethod}{Computational Thinking Model\xspace}
\makeatother

\def\BibTeX{{\rm B\kern-.05em{\sc i\kern-.025em b}\kern-.08em
    T\kern-.1667em\lower.7ex\hbox{E}\kern-.125emX}}
\begin{document}

\title{Computational Thinking Reasoning in Large Language Models
\thanks{\textsuperscript{$\dagger$}Ge Li and Zhi Jin are the corresponding authors.}\thanks{\textsuperscript{*}Work done during Kechi Zhang's internship at ByteDance.} }

\author{\IEEEauthorblockN{
Kechi Zhang\textsuperscript{1*},
Ge Li\textsuperscript{1$\dagger$},
Jia Li\textsuperscript{1},
Huangzhao Zhang\textsuperscript{1},
Jingjing Xu\textsuperscript{2}, \\
Hao Zhu\textsuperscript{1},
Lecheng Wang\textsuperscript{1},
Jia Li\textsuperscript{3},
Yihong Dong\textsuperscript{1},
Jing Mai\textsuperscript{1}, \\
Bin Gu\textsuperscript{4}, and
Zhi Jin\textsuperscript{1$\dagger$}
}
\IEEEauthorblockA{\textsuperscript{1}Key Lab of High Confidence Software Technology (PKU), Ministry of Education\\
School of Computer Science, Peking University, China}
\IEEEauthorblockA{\textsuperscript{2}ByteDance}
\IEEEauthorblockA{\textsuperscript{3}College of AI, Tsinghua University}
\IEEEauthorblockA{\textsuperscript{4}Beijing Institute of Control Engineering}
\IEEEauthorblockA{\texttt{\{zhangkechi,lige,zhijin\}@pku.edu.cn}}
}

\maketitle
\begin{abstract}

While large language models (LLMs) have demonstrated remarkable reasoning capabilities, they often struggle with complex tasks that require specific thinking paradigms, such as divide-and-conquer and procedural deduction, \etc
Previous researches integrate external, reliable tools to alleviate logical inconsistencies and hallucinations in LLMs' problem-solving processes.
However, we argue that the root challenge is more profound: LLMs lack the complex thinking paradigms (\ie, computational thinking) during reasoning.
In this paper, we propose \fullmethod (\method), a novel framework that incorporates computational thinking paradigms into LLMs. 
This framework enables LLMs to reformulate complex problems through decomposition, abstraction, reduction, and simulation, among other techniques.
Specifically, live code execution is seamlessly integrated into the reasoning process, allowing \method to think by computing.
\method directly instills computational thinking objectives into LLMs through tailored reinforcement learning rewards, which encourages problem simplification, modular planning, and iterative verification.
We conduct extensive evaluations on multiple code generation and mathematical benchmarks.
The results demonstrate that \method outperforms conventional reasoning models and tool-augmented baselines in terms of accuracy, interpretability, and generalizability.
We hope this study offers valuable insights for AI reasoning, where LLMs can transform problems into robust, verifiable, and scalable computational workflows, much like computer scientists do.

\end{abstract}


\section{Introduction}

Recent advancements in large language models (LLMs), particularly those enhanced through reinforcement learning (RL), have significantly pushed the boundaries of what these models can achieve in natural language reasoning \cite{openaio1}. 
Models such as DeepSeek-R1 \cite{guo2025deepseek} have demonstrated remarkable capabilities. However, in complex domains like algorithmic problem-solving and mathematical reasoning, these models often face fundamental limitations when dealing with structured and symbolic reasoning tasks, where precision, consistency, and compositionality are essential. 
In these contexts, errors in intermediate steps often propagate through the reasoning pipeline unchecked, making it challenging for the models to produce reliable solutions \cite{lightman2023let,chen2022program,hendrycks2021measuring, jain2024livecodebench,zhang2024codedpo,zhang2024deep,zhang2025focused}.

\textbf{A critical limitation arises from the fact that most natural language reasoning models perform reasoning as a purely generative process.} 
While advanced chain-of-thought prompting enables the articulation of intermediate reasoning steps, it does not provide the model with mechanisms to systematically verify its outputs or revise erroneous intermediate conclusions.
Some current researches alleviate these issues by supplementing the reasoning capabilities of LLMs with external and reliable tools, such as web search and code interpreter \cite{schick2023toolformer, feng2025retool, qian2025toolrl, li2025torl, openhands}.
Although these augmentations are beneficial, LLMs often treat reasoning as an opaque generation process.
In this context, the usage of tools serves as auxiliary support rather than being integrated into a cohesive thinking paradigm.
To address these challenges, we contend that the fundamental limitation is not merely the lack of tools, but rather the absence of essential thinking paradigms that underpin the reasoning process. 
For computer scientists, foundational methodologies such as decomposition, reduction, abstraction, and simulation form the core of computational thinking \cite{wing2006computational}, enabling them to tackle complex problems effectively.

In general, computational thinking is a cognitive paradigm that involves transforming complex problems into more manageable forms through methodologies such as decomposition, enumeration and verification.
\Eg, those who have mastered computational thinking often ask critical questions during their problem-solving process, including but not limited to the following questions \cite{wing2006computational}:
Can the given task be broken down into modular subgoals?
Can the current optimal state be derived from previous states?
The thinking paradigm underlying these questions fosters deep and systematic reasoning, enabling more effective problem-solving.

\begin{figure}[h]
    \centering
    \includegraphics[width=\linewidth]{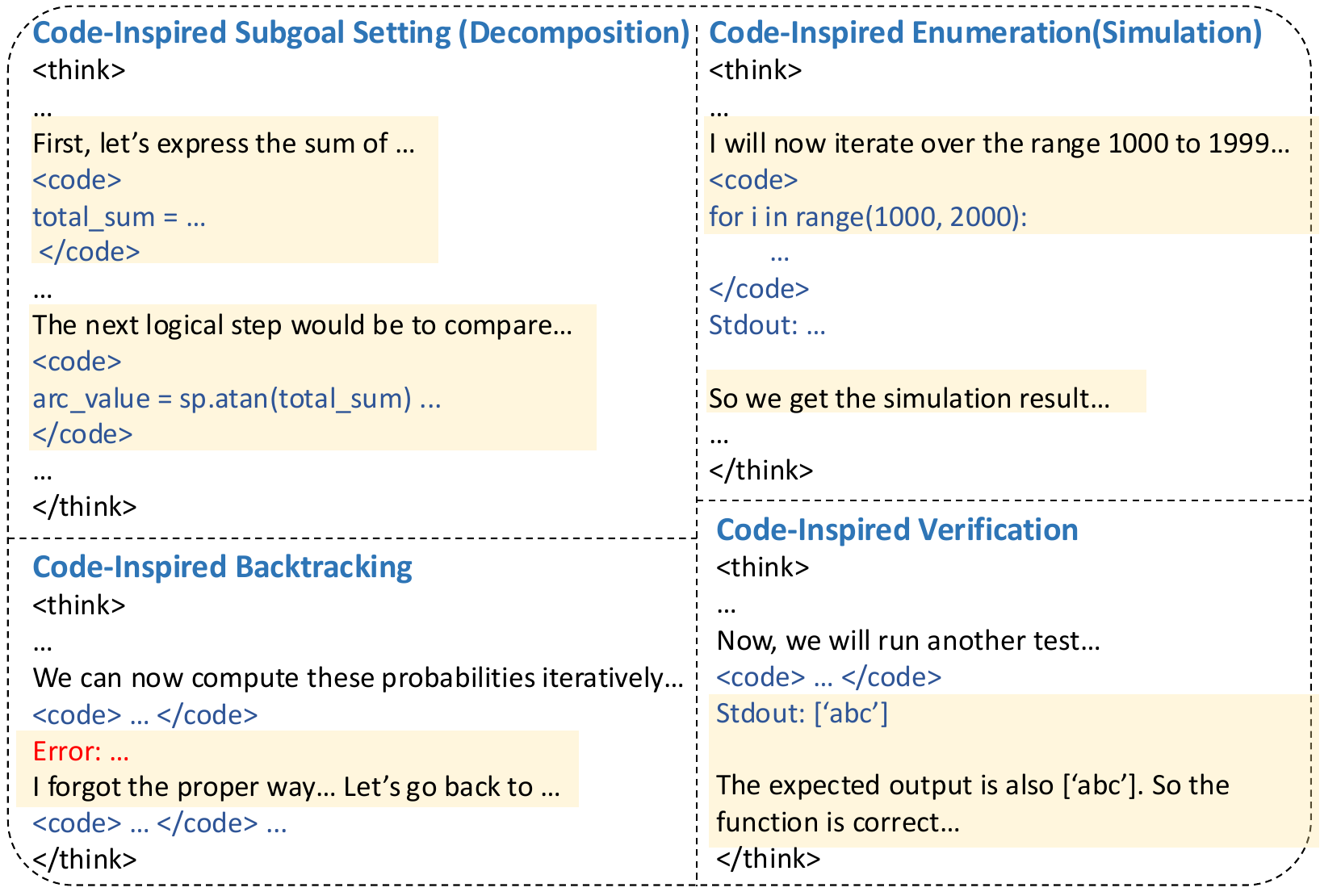}
    \caption{Computational thinking with code inspired patterns from our model.}
    \label{fig:introcase}
\end{figure}

Drawing inspiration from this principle, we propose \fullmethod (abbreviated as \method), a framework that enables LLMs to integrate computational thinking as a primary reasoning strategy. 
Specifically, \method provides an interactive execution environment where the reasoning process of LLMs is enhanced through live code execution. 
The multi-turn interaction facilitates decomposition of monolithic reasoning tasks, abstraction beyond textual limitations, and explicit intermediate computation in place of implicit speculation.
\method allows LLMs to learn computational thinking within this environment through reinforcement learning (RL), with supervised fine-tuning (SFT) serving as the cold start warmup.
LLMs acquire specific thinking methodologies from our constructed dataset during the SFT phase, and they further expand these capabilities during the RL phase. 
The resulting model exhibits behavior that is not only more accurate, but also systematically grounded in executable logic. As illustrated in Figure \ref{fig:introcase}, the incorporation of code-inspired computational thinking enables the model to solve complex reasoning tasks with great consistency, transparency, and precision.

We conduct extensive experiments on code and mathematical benchmarks \cite{chen2021evaluating,austin2021program,jain2024livecodebench,li2022competition,aime,lightman2023let}.
Experimental results suggest that \method outperforms conventional reasoning LLMs  as well as tool-augmented baselines \cite{guo2025deepseek, openaio1, ahmad2025opencodereasoning,claude,qwq32b,qwen2.5,openhands,Bespoke,openthinker,openr1}.
\method not only introduces a new technique but also offers a novel perspective on how LLMs can reason more effectively. 
By adopting the thinking paradigms of computer scientists, LLMs are better equipped to solve problems that require compositionality and precision.
The framework of \method holds great potential to make a broad impact on future intelligent systems.

The key contributions of this paper are listed as follow:

\begin{itemize}
    \item We argue that essential thinking paradigms are missing in LLMs. This absence underpins their reasoning processes and restricts their ability to solve complex problems.
    \item We propose integrating computational thinking into LLMs as primary reasoning methodologies, resulting in the framework of \method. With a live execution environment, \method enables LLMs to practice thinking paradigms such as code inspired verification during reasoning processes.
    \item We introduce a two-phase training strategy that includes SFT on constructed structured reasoning datasets and RL using a custom-designed reward function. This approach enables \method to autonomously learn adaptive reasoning strategies and behaviors that are aligned with problem-solving objectives.
    \item We demonstrate through comprehensive experiments across diverse benchmarks—including math-heavy reasoning tasks (\eg AIME) and code-based tasks (\eg LiveCodeBench)—that \method consistently outperforms state-of-the-art methods, achieving higher accuracy, interpretability, and generalizability.
\end{itemize}

\section{Motivating Examples}
\label{sec:motivating}
We present a motivating example to elaborate on the limitation issue within current natural language based reasoning models in Figure \ref{fig:motivation}.
Despite their great performance under algorithmic code generation and mathematics, these LLMs still struggle in several key points that affect the reasoning.

\subsection{Limitations in Natural Language-based Reasoning}
\subsubsection{The Rare "Aha Moment" with Self-Edit Capabilities}

\begin{figure}
    \centering
    \includegraphics[width=\linewidth]{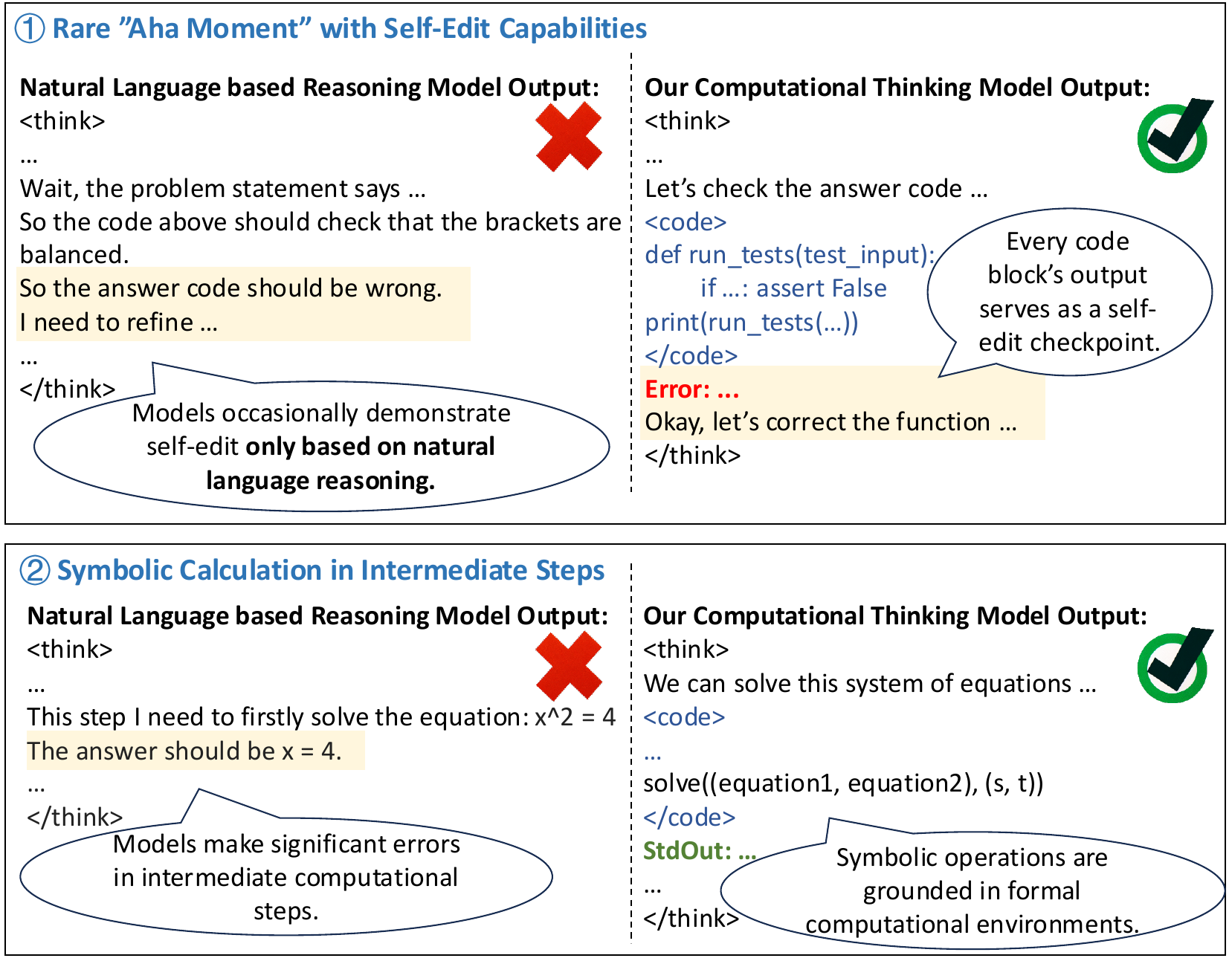}
    \caption{Limitation cases in natural language-based reasoning models. (Cases in the left part are from DeepSeek-R1, one of the most powerful reasoning models.)}
    \label{fig:motivation}
\end{figure}

Current language models demonstrate an intriguing phenomenon in their reasoning behavior. While capable of generating multi-step solutions through chain-of-thought prompting, they typically produce linear reasoning paths with limited capacity for self-correction. This creates a fundamental limitation where errors in early reasoning steps propagate unchecked through subsequent stages, as the models lack robust mechanisms to identify and revise flawed intermediate conclusions.

What makes this particularly interesting is that when models do occasionally demonstrate effective self-edit during training, researchers often describe it as an "aha moment" (see the left upper part of Figure \ref{fig:motivation}). 
These moments are noteworthy not because they represent paradoxical behavior, but because they reveal the model's latent potential for self-editing that is otherwise difficult to elicit. The rarity of these occurrences highlights how challenging it is to train models to consistently recognize and correct their own reasoning errors.

Human problem-solving naturally incorporates continuous self-monitoring and adjustment-a capability that remains largely absent in current AI systems. While reinforcement learning approaches have made progress by introducing external feedback loops, the models still struggle with intrinsic self-correction. The celebrated ``aha moments'' represent promising but isolated cases where models overcome this limitation, suggesting that more systematic approaches are needed to make self-correction a reliable rather than serendipitous capability.

This gap becomes particularly apparent in complex reasoning tasks, where the ability to dynamically revise one's approach could significantly improve performance. The current situation presents both a challenge and an opportunity: while spontaneous self-correction remains rare, its occasional occurrence proves that models do possess this capability in principle, waiting to be properly harnessed through improved training paradigms.

\subsubsection{Symbolic Calculation in Intermediate Steps}

Current LLMs approach symbolic operations (\eg algebraic manipulation, logical inference) through textual pattern matching rather than structured computation, as discussed in prior work \cite{hu2024case}. While these models can approximate symbolic transformations using statistical correlations from their training data, they frequently violate fundamental mathematical rules and semantic constraints. This limitation manifests clearly in tasks requiring rigorous symbolic manipulation, such as polynomial factorization or modular arithmetic, where models often generate syntactically plausible but mathematically invalid derivations.

The core issue stems from LLMs treating mathematical symbols as surface-level tokens rather than as manipulable objects governed by formal systems. Our analysis of the powerful DeepSeek-R1 model reveals this fundamental weakness: when presented with complex calculation tasks, the model not only produces incorrect final answers but also makes significant errors in intermediate computational steps (see the left bottom part of Figure \ref{fig:motivation}). Notably, these errors occur even for simple calculations. This observation highlights the inherent risks of relying solely on natural language-based reasoning for symbolic computation, regardless of problem complexity. The persistent failure to maintain mathematical validity across both simple and complex operations suggests a fundamental limitation in the current paradigm's ability to support rigorous symbolic reasoning.

\subsection{Computational Thinking as a Foundational Remedy}
As \cite{wing2006computational} says, \textit{computational thinking involves solving problems, designing systems, and understanding human behavior, by drawing on the concepts fundamental to computer science. Computational thinking includes a range of mental tools that reflect the breadth of the field of computer science.}
Our Computational Thinking Model (CTM) addresses these limitations by reconceptualizing original occasional reasoning as an iterative, executable process. 
By interleaving natural language with live code execution, CTM introduces execution feedback loops that enable dynamic self-edit—every code block’s output serves as a validation checkpoint in Figure \ref{fig:motivation}. Symbolic operations are grounded in formal computational environments, ensuring rule-based rigor. 
% CTM can also well handle the long dependency within the execution environment in the generated output which is also an essential point for LLMs \cite{zhang2024hirope}.
This paradigm shift from generative text construction to computational workflow execution yields more reliable, scalable, and verifiable problem-solving.
It helps our CTM show various complex code inspired reasoning behaviors, as shown in Figure \ref{fig:introcase}.

\section{Method}
\label{sec:method}

\begin{figure}
    \centering
    \includegraphics[width=\linewidth]{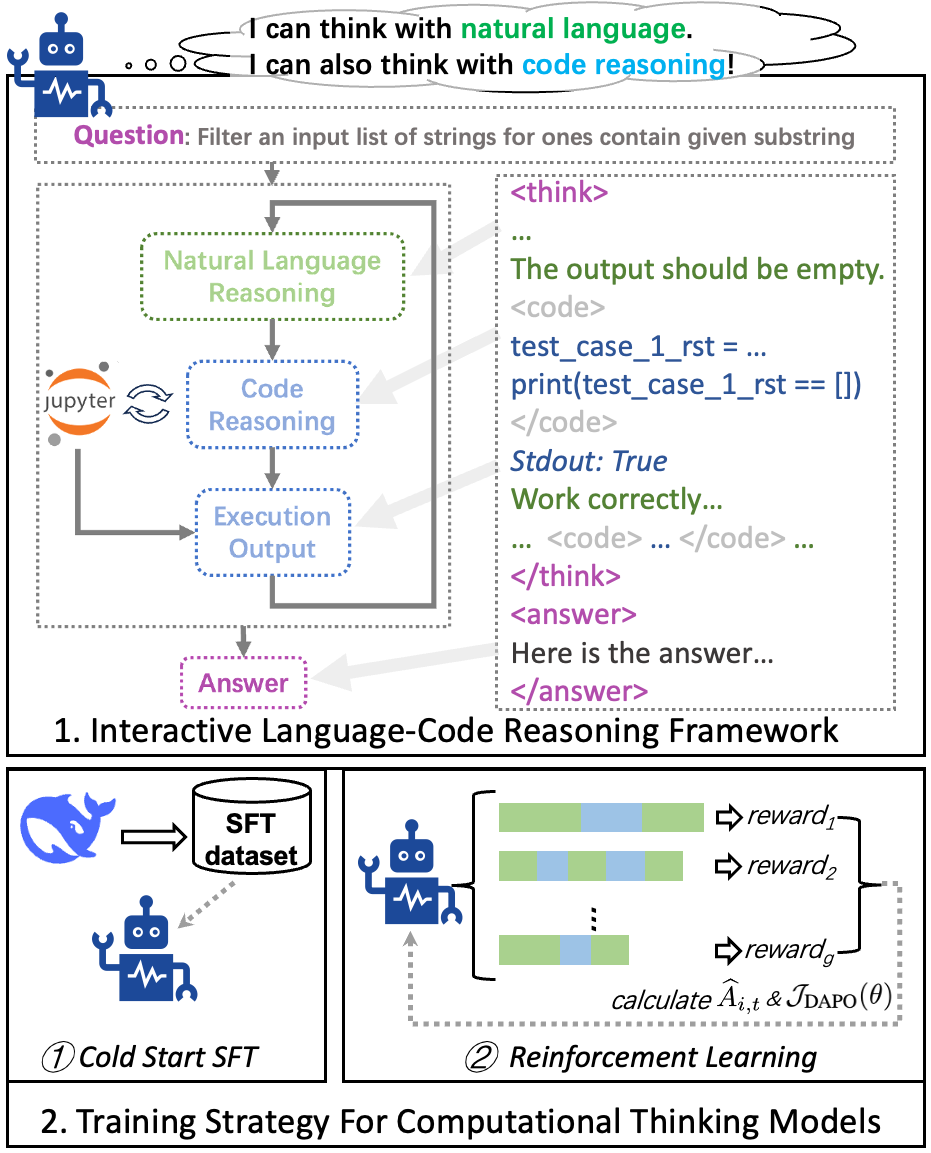}
    \caption{The illustration of the framework and training strategy for our computational thinking model.}
    \label{fig:ctm_pipeline}
\end{figure}

We propose the \textbf{Computational Thinking Model (CTM)}, an interactive reasoning framework that interleaves natural language analysis and code-based execution to tackle challenging tasks. In our computational thinking model, large language models alternate between planning/reflection and executable code, thereby grounding intermediate reasoning in concrete computations. 
In this section, we will describe our framework architecture, the training approach, and the inference procedure supported by real-time code execution in detail.

\subsection{Interactive Language-Code Reasoning Framework}
\label{sec:method_framework}

The core idea behind our computational thinking model is to let the model ``think by computing.'' 
Rather than a purely text-based chain-of-thought, our model intersperses generated Python code blocks with self-reflective language instructions, all stored in a shared conversational context. 
Figure \ref{fig:ctm_pipeline} illustrates how the model iterates between natural language reasoning, code reasoning, and execution feedback.

\subsubsection{Code Execution Integration in Computational Thinking Models}

Starting from an initial input question or prompt, the Computational Thinking Model (CTM) iteratively appends each newly generated natural language reasoning or code snippet to the shared reasoning context. Whenever the model emits a code block enclosed within \texttt{<code>...</code>} tags, the system extracts the enclosed code snippet and executes it in a sandbox environment with persistent memory—functionally similar to the notebook-style execution used in environments like Jupyter. This execution mechanism enables the model to iteratively validate hypotheses, perform computations, and refine intermediate steps, much like a human programmer leveraging an interactive coding platform to test and debug solutions during problem-solving tasks.

The results of code execution, including standard outputs (\texttt{stdout}), error messages (\eg syntax or runtime errors), or returned values, are automatically captured and transformed into text. These textual outputs are subsequently appended to the model's reasoning context, serving as feedback for the next iteration. This structured feedback loop not only grounds the model's reasoning in verifiable computation but also enables real-time debugging and refinement, facilitating dynamic problem-solving. The iterative process continues until the model generates an explicit \texttt{} tag, signaling the completion of the reasoning process and providing the final solution.

This loop mirrors the behavior of human programmers who incrementally solve problems by writing, executing, and refining code within interactive computing environments. It allows CTM to model the problem-solving process with a higher degree of flexibility, accuracy, and interpretability. The underlying process is outlined in Algorithm \ref{alg:ctm_interactive}, which illustrates the interleaving of reasoning, code generation, execution, and contextual feedback.

\begin{algorithm}[t]
\caption{Interactive Language and Code Mixed Reasoning in CTM}
\label{alg:ctm_interactive}
\begin{algorithmic}[1]
\Procedure{InteractiveCTM}{$Q$: input question}
    \State $H \gets [\textbf{initial prompt: } Q]$  \Comment{Start with \texttt{<think>}}
    \While{\texttt{</think>} not in $H$}
        \State $o \gets \Call{LLM}{ H }$  \Comment{Model generates next step until get \texttt{</code>}}
        \If{\texttt{<code>} in $o$}
            \State $\texttt{code} \gets$ extract code block from $o$
            \State $\texttt{result} \gets \Call{RunInSandbox}{\texttt{code}}$
            \State $H \gets H + (o + \texttt{</code>} + \texttt{result})$
        \Else
            \State $H \gets H + o$
        \EndIf
    \EndWhile
    \State $ans \gets \Call{LLM}{ H }$  \Comment{Start with \texttt{<answer>} until get \texttt{</answer>}}
    \State \Return answer enclosed by \texttt{<answer>}...\texttt{</answer>}
\EndProcedure
\end{algorithmic}
\end{algorithm}

This algorithm outlines the system's interactive reasoning process, where the model alternates between generating reasoning steps, executing code, and integrating execution results as feedback. The sandbox environment is particularly crucial, as it supports persistent memory across code executions, allowing variable states, functions, and intermediate results to persist across iterations. This feature enables the model to build upon its own computations dynamically, reflecting the iterative and modular nature of computational thinking.

By embedding computational feedback directly into the reasoning process, CTM transcends traditional static reasoning paradigms and achieves a level of adaptability and transparency akin to human problem-solving strategies.

\subsubsection{Execution Environment and Result Integration}

We implement a notebook-style \emph{execution environment} that persists state across multiple code cells. 
This environment can capture syntax or runtime errors and prints them as textual feedback, records standard outputs, and makes newly defined variables or functions accessible to subsequent code blocks.  
Hence, real-time \emph{compute-and-check} stages guide the model’s reasoning by enabling dynamic refinement: if a mistake arises in part of the solution, the model can revise its code and retest until reaching a sufficiently correct final answer.

In practice, to balance training efficiency and effectiveness, we seek a stable, accurate, and responsive code interpreter implementation.
We choose \textit{Sandbox Fusion} \cite{sandbox}, and use its provided \textit{Jupyter} mode for our model, which provides an isolated execution environment. Despite slightly higher latency, it delivers superior stability for sustained training
operations.
We implemented specific error handling optimizations to enhance training effectiveness. When \textit{Sandbox Fusion} gives out execution errors, it generates verbose tracebacks containing irrelevant file path information. To reduce context length and preserve only relevant error information, we extract only the final line of error messages.

\subsection{Training Strategy For Computational Thinking Models}
\label{sec:method_training}

To effectively incorporate computational thinking principles into our model's reasoning capabilities, we employ a two-phase training strategy designed to align the model’s reasoning with natural language and code execution. These phases consist of:
(1) an initial supervised fine-tuning (``cold start'') phase to familiarize the model with natural language and code-mixed reasoning traces, and
(2) a reinforcement learning (RL) phase leveraging code-based rollouts and Decouple Clip and Dynamic Sampling Policy Optimization (DAPO) \cite{yu2025dapo}, an advanced policy optimization algorithm inspired by GRPO \cite{shao2024deepseekmath}.
Below, we describe each phase in detail.

\subsubsection{Cold Start Supervised Fine-tuning}
\label{sec:sft}

The supervised fine-tuning (SFT) phase aims to establish the foundational reasoning patterns required for integrating computational thinking into the model. To achieve this, we trained the model on curated datasets comprising reasoning tasks annotated with mixed natural language and executable code traces. These traces are designed to interleave natural language explanations and code snippets while reflecting the iterative problem-solving process characteristic of computational thinking.

The training datasets include a diverse range of tasks drawn from both mathematical reasoning and programming challenges.
We sample 13k code questions and 11.9k math questions from existing open-sourced dataset \cite{ahmad2025opencodereasoning, areal}.
We use the few-shot prompt to guide advanced language models like DeepSeek-V3 and DeepSeek-R1 \cite{guo2025deepseek,liu2024deepseek} to generate natural language and code mixed solution traces.
Each solution trace consists of a sequence of structured reasoning steps, including hypothesizing, implementation attempts, debugging, refinement, and validation. For example, a typical trace might begin with a high-level hypothesis, followed by a naive implementation, iterative corrections based on execution feedback, and finally an optimized solution validated against the problem constraints.
The training corpus is designed to encourage the model to adopt structured reasoning patterns that align closely with computational thinking. These patterns may involve strategies like divide-and-conquer decomposition, dynamic programming, heuristic search, or brute-force refinement. In practice, we construct some few-shot examples in these patterns to help the dataset construction.
For instance, a prototypical trace might follow the workflow: \texttt{understand} → \texttt{plan} → \texttt{code} → \texttt{validate} → \texttt{refine} → \texttt{finalize}. 
By exposing the model to such structured reasoning patterns, the fine-tuning phase effectively establishes computational thinking concepts—including decomposition, abstraction, iteration, simulation, and transformation—as integral components of the reasoning process.

During SFT training, we mask out execution outputs from the sandbox environment (\eg runtime results or error messages). This ensures that the model does not rely on predicting specific values or computational outputs directly but instead uses reasoning patterns to arrive at consistent and generalizable solutions. Such masking prevents over-reliance on data memorization and significantly improves training stability by guiding the model to focus on logical patterns rather than specific numerical outcomes. For instance, instead of inferring execution outputs in purely natural language contexts, the model learns to scaffold its reasoning around the structural process of problem-solving.
Furthermore, SFT encourages the model to leverage computational feedback during intermediate stages, laying the foundation for dynamic refinement and self-correction. This flexibility enables the model to adapt to diverse problem-solving scenarios without being constrained by rigid reasoning templates, making it highly versatile across tasks requiring structured yet adaptive reasoning.

\subsubsection{Reinforcement Learning with NL-Code-Mixed Rollouts}
\label{sec:rl}

Following the supervised fine-tuning phase, we further align the model's behavior by employing reinforcement learning (RL). This phase optimizes the model’s reasoning efficiency and reliability across both natural language and executable code. By generating dynamically mixed rollouts of natural language and code snippets, the model develops emergent strategies characteristic of computational thinking, further refining its ability to reason iteratively and adaptively.

To ensure stable optimization in the presence of mixed modality outputs, we adopt the DAPO algorithm \cite{yu2025dapo}, a tailored variant of GRPO. The RL training process begins by sampling multiple reasoning trajectories (or “rollouts”) for each task and evaluating them against a rule-based reward function. We sample 2k code questions and 2k math questions from open-sourced dataset \cite{skyworkrldata}.

\paragraph{Reward Design and Trajectory Rollout Sampling}
Given a question $q$ with a ground-truth answer $a$, the model samples $G$ reasoning trajectories $\{o_i\}_{i=1}^G$ from the current policy $\pi_{\theta_\text{old}}$, where each trajectory interleaves natural language reasoning steps and executable \texttt{<code>...</code>} blocks within the shared sandbox environment. To provide supervision, each trajectory is assigned a scalar reward $R_i$, reflecting the quality of its final output and intermediate reasoning. Rewards are computed via domain-specific evaluation tools as follows:

\begin{itemize}
    \item \textbf{Code Question Judgments:} For programming tasks, correctness is determined by the solution's ability to pass all test cases within the sandbox environment, as implemented by \emph{Sandbox Fusion} \cite{sandbox}. Correct solutions are assigned a reward of $+1$, while solutions that fail the test cases receive $-1$. Execution errors (\eg syntax or runtime issues) are also penalized during trajectory evaluation.
    \item \textbf{Mathematical Question Judgments:} For mathematical reasoning tasks, correctness is determined using the \emph{math-verify} tool \cite{mathverify}, which checks for exact matches or equivalence to the ground-truth answer. Accurate solutions receive a reward of $+1$, while incorrect solutions are assigned $-1$.
\end{itemize}

Each trajectory incorporates both the final answer and the natural language/code reasoning trace generated during the process, ensuring that intermediate steps contribute indirectly to the optimization.

\paragraph{Reinforcement Learning Objective with DAPO}
To align the model’s policy $\pi_\theta$ with computational thinking principles, DAPO maximizes the normalized token-level advantage while constraining updates using asymmetric clipping to ensure stable learning. For each sampled question-answer pair $(q, a)$ and its associated rollouts $\{o_i\}_{i=1}^G$, the optimization objective is defined as:

\begin{equation}
\label{eq:dapo}
\mathcal{J}_{\text{DAPO}}(\theta)
= \mathbb{E}_{(q,a), \{o_i\}} \Bigg[
   \frac{1}{\sum_{i=1}^G |o_i|}
   \sum_{i=1}^G \sum_{t=1}^{|o_i|}
   \min \bigl( r_{i,t}(\theta) \cdot \widehat{A}_{i,t}, \bigr.
\end{equation}
\[
   \Bigl. \text{clip}\bigl(r_{i,t}(\theta), 1-\epsilon_\text{low}, 1+\epsilon_\text{high}\bigr) \cdot \widehat{A}_{i,t} \bigr)
\Bigg].
\]
where $r_{i,t}(\theta) = \dfrac{\pi_\theta(o_{i,t} \mid q, o_{i,<t})}{\pi_{\text{old}}(o_{i,t} \mid q, o_{i,<t})}$ and $\widehat{A}_{i,t} = \dfrac{R_i - \mu_R}{\sigma_R}$. Only batches containing both successful and failed rollouts (\ie, $0 < \#\{\text{correct }o_i\} < G$) are used, ensuring a meaningful gradient signal.
This training approach drives the model to iteratively refine complex solutions by writing code, evaluating the results, and adjusting generation patterns accordingly, reinforcing the “computational thinking” principle.

\subsection{Inference Enhanced with Code Execution Environment}
\label{sec:method_infer}

At inference time, CTM applies the same language-code reasoning loop as during training. Given an input query, the model alternates between generating natural language reasoning steps and executable code snippets. Each code block is executed in real-time within a sandbox environment, and the generated output—whether standard print responses, errors, or returned values—is appended to the reasoning history. The model dynamically adjusts its reasoning trajectory in response to execution outcomes, iteratively refining its solution until a final answer is produced. This iterative refinement ensures that errors detected during intermediate steps can be corrected, reducing the likelihood of logical or computational failures.

Compared to original natural language based reasoning models, CTM's iterative language-code cycle enhances robustness and interpretability, as each step in the reasoning process is grounded by verifiable computations. This approach not only facilitates accurate problem-solving but also provides a transparent trace of the reasoning process, making CTM a versatile framework for tasks requiring multi-step reasoning and computational rigor.

\section{Study Design}

\subsection{Research Questions}

To evaluate the effectiveness of our CTM, we conduct a comprehensive large-scale study. We aim to answer the following research questions (RQs):

\textbf{RQ1: How does CTM perform on code-related problem-solving benchmarks?}
We evaluate our model on six real-world programming benchmarks (\ie  LiveCodeBench \cite{jain2024livecodebench}, CodeContests \cite{li2022competition}, HumanEval \cite{chen2021evaluating}, HumanEval+ \cite{liu2023your}, MBPP \cite{austin2021program}, and MBPP+ \cite{liu2023your}), from function-level code generation to competition-level coding challenges. The goal is to demonstrate whether CTM  excels in producing correct code by leveraging its inherent computational thinking capabilities.

\textbf{RQ2: What is the performance of CTM on other general-domain reasoning benchmarks like mathematics?}
To evaluate the generalizability of our model beyond code tasks, we also assess its performance on reasoning-intensive mathematical benchmarks, including AIME 2024, AIME 2025 \cite{aime}, and MATH-500 \cite{lightman2023let}. These benchmarks encompass both elementary and advanced problem-solving tasks that require complex reasoning and precise intermediate computation.

\textbf{RQ3: How does CTM influence the reasoning trajectories of LLMs?}
To gain deeper insights into the cognitive process of LLMs equipped with computational thinking, we investigate their reasoning trajectories—examining how problems are decomposed into smaller subproblems, how intermediate solutions are computed, and how errors are corrected \cite{gandhi2025cognitive,zeng2025simplerl}. This analysis emphasizes whether CTM's reasoning behavior aligns with computational thinking principles like decomposition, abstraction, and iteration.

\textbf{RQ4: What is the impact of design choices within CTM?}
Design choices during training might significantly influence the model’s effectiveness. To evaluate this, we conduct extensive ablation studies that vary key components, such as (1) reward design during reinforcement learning, (2) curriculum in training datasets (\eg code vs. general tasks), and (3) the use of interactive mixed-language environments (\eg incorporating Jupyter-style execution). This analysis can effectively reflect crucial factors that contribute to our framework.

\subsection{Benchmarks and Metrics}

Our study is motivated by the principle of Software Engineering for AI (SE4AI), borrowing concepts from computational thinking to address diverse reasoning-intensive problems, including both code-related tasks (\eg code generation) and other general reasoning tasks like mathematics. In this paper, we evaluate CTM on nine benchmarks that cover a broad spectrum of reasoning challenges, encompassing both code and non-code domains.

For evaluating model performance on code-related tasks, we introduce six widely used programming benchmarks, including both functional code generation tasks and programming contest tasks. Functional code generation benchmarks contain HumanEval, HumanEval+, MBPP, and MBPP+ \cite{chen2021evaluating,austin2021program,liu2023your}, focusing on functional correctness in code generation with extended test cases. Programming contest benchmarks include CodeContests and LiveCodebench \cite{li2022competition,jain2024livecodebench}, where CodeContests has 165 programming code contest questions from diverse platforms, and LiveCodebench has the newest contest questions released between May 2023 and January 2025, containing 880 problems.

For further evaluating reasoning capabilities beyond programming, we apply three benchmarks in mathematical problem-solving. AIME 2024 and AIME 2025 benchmarks \cite{aime} contain problems from the American Invitational Mathematics Examination. These problems are prestigious high school mathematics competition problems known for its challenging mathematical problems.
MATH-500 \cite{lightman2023let} is collected from real-world math problems verified by OpenAI, which requires long-horizon reasoning.

To evaluate the performance of CTM, we execute the generated code on the hidden test cases to get the pass rate for each code question. For mathematical tasks, we use the \textit{math-verify} \cite{mathverify} to evaluate the generated output and the ground truth output to get the accuracy for math questions.

\subsection{Compared Baselines}

We evaluate our approach against three categories of state-of-the-art reasoning systems to provide comprehensive performance comparisons, including proprietary source models, open-source models, and tool-augmented agents.

\textbf{Proprietary Source Models}: We compare CTM against the advanced proprietary source models, containing Anthropic's \texttt{Claude-3.5-Sonnet} \cite{claude}, \texttt{OpenAI's GPT-4o-0513} \cite{gpt4o}, and \texttt{o1} series models \cite{openaio1}. They represent the current pinnacle of commercial reasoning capabilities, with some results drawn from official benchmarks where available.

\textbf{Open-Source Models}: These models in our comparison comprise several open-source systems optimized specifically for programming tasks, such as \texttt{Bespoke-Stratos-32B} \cite{Bespoke}, \texttt{OpenThinker-32B} \cite{openthinker}, \texttt{OlympicCoder-32B} \cite{openr1}, and \texttt{OCR-Qwen-32B} \cite{ahmad2025opencodereasoning}, along with \texttt{DeepSeek-V3} \cite{liu2024deepseek} and \texttt{R1} \cite{guo2025deepseek}. While these models demonstrate strong performance on coding tasks through intensive domain-specific training, they typically show limitations when applied to broader mathematical reasoning problems. 

\textbf{Tool-Augmented Agents}: We also contrast our model with tool-augmented agent frameworks, where \texttt{OpenHands} \cite{openhands} is a representative example. It enhances LLMs' reasoning ability through integrated development environments that combine text editing, code execution, and multi-step planning capabilities. In this paper, we use the advanced \texttt{DeepSeek-V3} as the base LLM.

Considering that CTM is continuously trained from \texttt{Qwen2.5-32B-Instruct}, we evaluate the performance of \texttt{Qwen2.5-32B-Instruct} \cite{qwen2.5} to further analyze the effectiveness of our CTM. All comparisons maintain identical evaluation settings to ensure fair and consistent measurements.

\subsection{Experimental Setups}

In the training process, we initialize CTM with \texttt{Qwen-2.5-32B-Instruct} \cite{qwen2.5}, where we fine-tune it under supervised fine-tuning (SFT) on computational thinking-flavored datasets (\S \ref{sec:sft}) and RL-based fine-tuning for more effective problem-solving behaviors (\S \ref{sec:rl}). 
In this paper, all RL-based experiments are conducted using the \textit{veRL} framework, adopting the DAPO algorithm for RL. 
We sample a batch size of 32 queries per training step, generating 16 responses per query. 
We utilize a Jupyter-style interactive execution environment. This environment allows the model to generate, execute, and iteratively refine code in real-time. We ultimately selected Sandbox Fusion \cite{sandbox}, which provides an isolated execution environment. Despite slightly higher latency, it delivers superior stability for sustained training
operations.
All experiments are conducted on 32 A100-80GB GPUs. 
Training typically completes within one week for our experiments under the above configurations.

In the inference process, greedy search is adopted as it ensures deterministic and consistent results for evaluation. Some evaluation results are referenced from their official reports \cite{yang2025qwen3, guo2025deepseek, ahmad2025opencodereasoning}.

\section{Experimental Results}

In this section, we present the experimental results to evaluate the performance of our CTM across various benchmarks, addressing the research questions outlined earlier. 

\begin{table}[t]
\centering
\caption{Code Generation Performance on Complex Competitive Programming Tasks}
\label{tab:sota_livecode_codecontests}
\begin{tabular}{lcc}
\toprule
\textbf{Model} & \textbf{LiveCodeBench} & \textbf{CodeContests} \\
\midrule
\textbf{Proprietary Source Models} \\
Claude-3.5-Sonnet-1022 & 38.9 & - \\
GPT-4o-0513 & 32.9 & - \\
OpenAI o1-mini & 53.8 & - \\
OpenAI o1-1217 & 63.4 & - \\
\midrule
\textbf{Open-Source Models} \\
DeepSeek-V3-671B & 36.2 & 20.2 \\
DeepSeek-R1-671B & 65.9 & 26.2 \\
QwQ-32B & 61.3 & 20.2 \\
Bespoke-Stratos-32B & 30.1 & 6.3 \\
OpenThinker-32B & 54.1 & 16.4 \\
OlympicCoder-32B & 57.4 & 18.0 \\
OCR-Qwen-32B & 61.8 & 24.6 \\
\midrule
\multicolumn{3}{l}{\textbf{Tool-augmented Agent Frameworks}} \\
OpenHands (DeepSeek-V3-671B) & 39.3 & 23.9 \\
\midrule
\textbf{Our Base Model} \\
Qwen2.5-32B-ins & 33.4 & 11.1 \\
\midrule
\textbf{Our Method} \\
CTM-32B-SFT & 61.6 & 24.2 \\
CTM-32B-RL & \textbf{66.5} & \textbf{28.5} \\
\bottomrule
\end{tabular}
\end{table}

\begin{table}[t]
\centering
\caption{Code Generation Performance on Function-level Code Generation Tasks}
\label{tab:sota_humaneval_mbpp}
\begin{tabular}{l@{}cccc}
\toprule
\textbf{Model} & \textbf{HumanEval} & \textbf{H+} & \textbf{MBPP} & \textbf{M+} \\
\midrule
\multicolumn{5}{l}{\textbf{Proprietary Source Models}} \\
Claude-3.5-Sonnet-1023 & 92.1 & 85.4 & 86.0 & 72.2 \\
GPT-4o-0513 & 91.5 & 87.8 & 89.4 & 74.6 \\
\midrule
\multicolumn{5}{l}{\textbf{Open-Source Models}} \\
DeepSeek-V3-671B & 94.5 & 89.0 & 90.2 & 74.9 \\
Bespoke-Stratos-32B & 88.4 & 84.1 & 87.6 & 74.1 \\
OpenThinker-32B & 88.4 & 85.4 & 89.4 & 75.4 \\
OlympicCoder-32B & 82.3 & 75.0 & 83.6 & 72.0 \\
OCR-Qwen-32B & 81.1 & 77.4 & 82.8 & 68.8 \\
\midrule
\multicolumn{5}{l}{\textbf{Tool-augmented Agent Frameworks}} \\
OpenHands (DS-V3-671B) & \textbf{95.5} & 90.5 & 91.0 & 76.6 \\
\midrule
\multicolumn{5}{l}{\textbf{Our Base Model}} \\
Qwen2.5-32B-ins & 88.4 & 83.7 & 84.0 & 74.3 \\
\midrule
\multicolumn{5}{l}{\textbf{Our Method}} \\
CTM-32B-SFT & 91.3 & 85.0 & 90.2 & 78.3 \\
CTM-32B-RL & \textbf{93.7} & \textbf{92.2} & \textbf{92.5} & \textbf{79.1} \\
\bottomrule
\end{tabular}
\end{table}

\begin{table}[t]
\centering
\caption{Math Performance Comparison with State-of-the-Art Models}
\label{tab:math}
\begin{tabular}{l@{}ccc}
\toprule
\textbf{Model} & \textbf{AIME 2024} & \textbf{AIME 2025} & \textbf{MATH-500} \\ 
\midrule
\textbf{Proprietary Source Models} \\
Claude-3.5-Sonnet-1022 & 16.0 & 13.3 & 78.3 \\
GPT-4o-0513 & 9.3 & 6.6 & 74.6 \\
OpenAI o1-mini & 63.6 & 60.0 & 90.0 \\
OpenAI o1-1217 & 79.2 & 73.3 & 96.4 \\
\midrule
\textbf{Open-Source Models} \\
DeepSeek-V3-671B & 39.2 & 36.6 & 90.2 \\
DeepSeek-R1-671B & 70.0 & \textbf{76.6} & \textbf{97.3} \\
QwQ-32B & \textbf{79.5} & 69.5 & 96.0 \\
Bespoke-Stratos-32B & 30.0 & 23.3 & 93.0 \\
OpenThinker-32B & 33.3 & 23.3 & 90.6 \\
OlympicCoder-32B & 40.0 & 36.6  & 94.0 \\
OCR-Qwen-32B & 36.6 & 30.0 & 92.2 \\
\midrule
\multicolumn{4}{l}{\textbf{Tool-augmented Agent Frameworks}}  \\
OpenHands (DS-V3-671B) & 50.0 & 46.6 & 93.6 \\
\midrule
\textbf{Our Base Model} \\
Qwen2.5-32B-ins & 6.6 & 6.6 & 86.6 \\
\midrule
\textbf{Our Method} \\
CTM-32B-SFT & 70.0 & 56.6 & 92.2 \\
CTM-32B-RL & \textbf{76.6} & \textbf{66.6} & \textbf{96.8} \\
\bottomrule
\end{tabular}
\end{table}

\subsection{RQ1: Performance on Code Problems} 
To answer RQ1, we first compare the results of our computational thinking models with the baseline LLMs and frameworks.
Tables \ref{tab:sota_livecode_codecontests} and \ref{tab:sota_humaneval_mbpp} summarize the performance of our model against state-of-the-art large reasoning models, code-specialized models, and agent frameworks on code-related benchmarks.
For function-level code generation tasks such as HumanEval(+) and MBPP(+), our model gains advantages from the code execution and refinement loop in our natural language and code-mixed reasoning traces. 
It can autonomously learn to correct errors that occur in the code and provide the final answer, as shown in Figures \ref{fig:motivation} and \ref{fig:casestudy}.
For complex code competition problems, our model demonstrates superior performance on both LiveCodeBench and CodeContests, two challenging benchmarks evaluating program reasoning and generation. 
Specifically, our method with reinforcement learning (RL) achieves the best results, scoring 66.5 on LiveCodeBench and 28.5 on CodeContests, outperforming other leading reasoning models such as \texttt{DeepSeek-R1} and code-specialized models like \texttt{OCR-Qwen-32B}. 
Compared to \texttt{OpenAI o1-mini} and \texttt{DeepSeek R1}, which are widely regarded as the state-of-the-art large reasoning models, our approach achieves significant gains with its interactive natural language and code mixed reasoning capabilities.

\subsection{RQ2: Performance on Math Problems}
Our model also achieves top-tier performance across those general domain tasks such as mathematical, as shown in Table \ref{tab:math}. 
Our RL-trained variant achieves results superior to or comparable with those of powerful proprietary and open-source models.
Specifically, our RL-trained model achieves scores of 76.6 on AIME 2024 and 66.6 on AIME 2025—complex competitive math benchmarks—and reaches 96.8 on the MATH-500 dataset. This performance demonstrates its capability to handle advanced symbolic reasoning and execute precise computations.
These results confirm that computational thinking as a paradigm induces natural language and code mixed reasoning behaviors that generalize well across code-specific and other general-domain tasks, making it highly versatile for reasoning-intensive applications.

To measure the contribution of our proposed training methodology, we further analyze the performance improvement for difference stages in our training process, relative to our base model (\texttt{Qwen2.5-32B-instruct}) which uses the same architecture but without computational thinking integration. Results are shown in Table \ref{tab:ablation}.
Our fine-tuned model (\texttt{SFT}) achieves substantial improvements across all benchmarks, highlighting the impact of task-specific supervised fine-tuning with computational thinking-flavored data.
Training with reinforcement learning further improves performance, with our \texttt{RL} variant achieving even larger performance boosts. 
Notably, it provides a gain of \textit{+33.1} on LiveCodeBench and \textit{+70.0} on AIME 2024 compared to the base model.
This demonstrates that the integration of computational thinking principles, both in supervised fine-tuning (SFT) and reinforcement learning (RL), significantly enhances the reasoning and problem-solving capabilities of LLMs.

\subsection{RQ3: Reasoning Behavior in Our CTM}
\label{sec:behavior}

\begin{figure}[htbp]
    \centering
    % \begin{subfigure}[b]{\linewidth}
        \includegraphics[width=\linewidth]{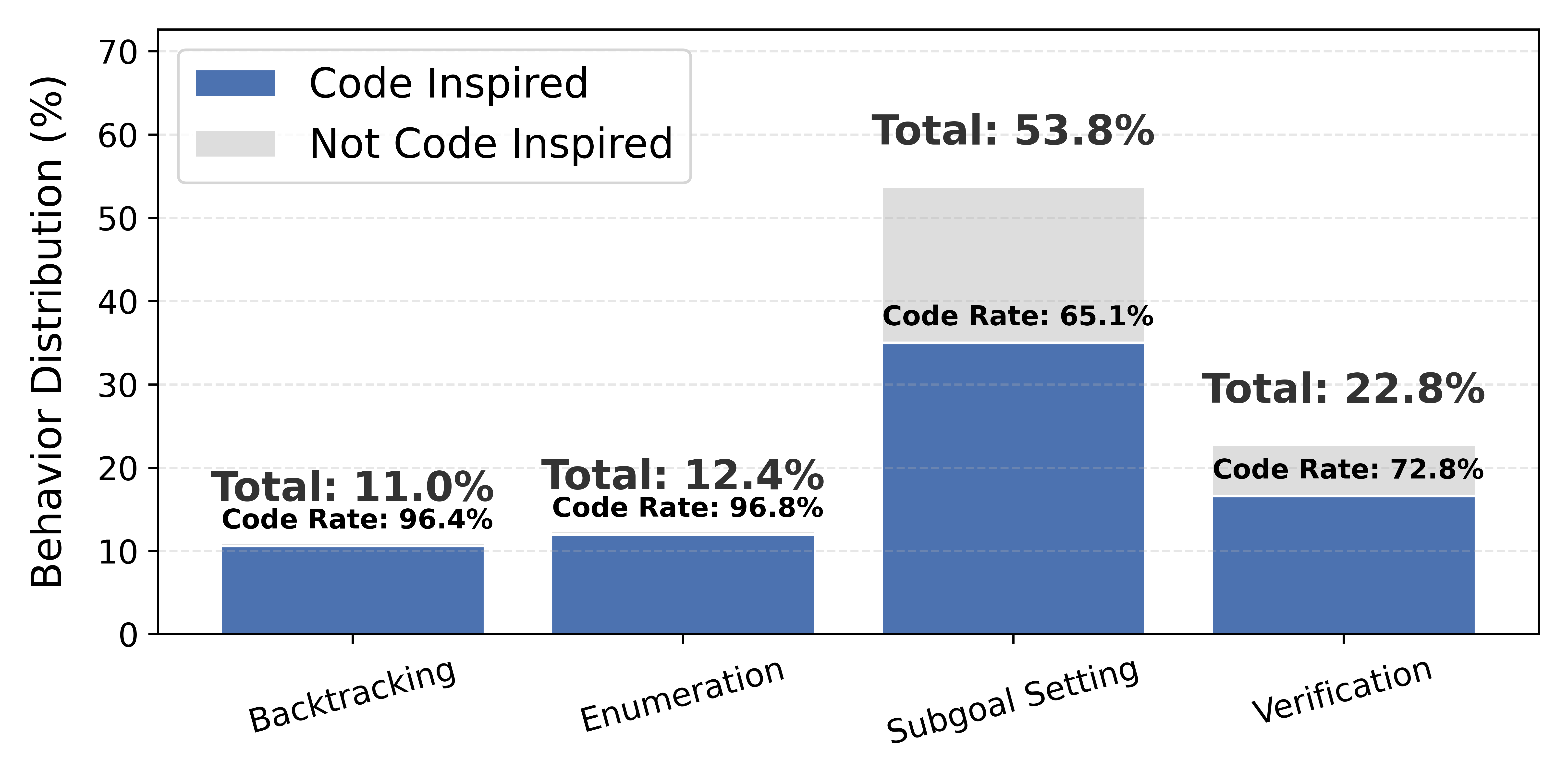}
        % \caption{Our Computational Thinking Model}
        % \label{fig:ctm_behav}
    % \end{subfigure}
    % \hfill
    % \begin{subfigure}[b]{\linewidth}
        % \includegraphics[width=\linewidth]{deepseekr1_behav.png}
        % \caption{DeepSeek-R1 Model}
        % \label{fig:deepseek_behav}
    % \end{subfigure}
    \caption{Comparison of Reasoning Behavior Distributions for our computational thinking model. Case studies for each behavior are shown in Figure \ref{fig:introcase}. }
    \label{fig:behavior_comparison}
\end{figure}

\begin{figure}[htbp]
    \centering
    \begin{subfigure}[b]{0.53\linewidth}
        \includegraphics[width=\linewidth]{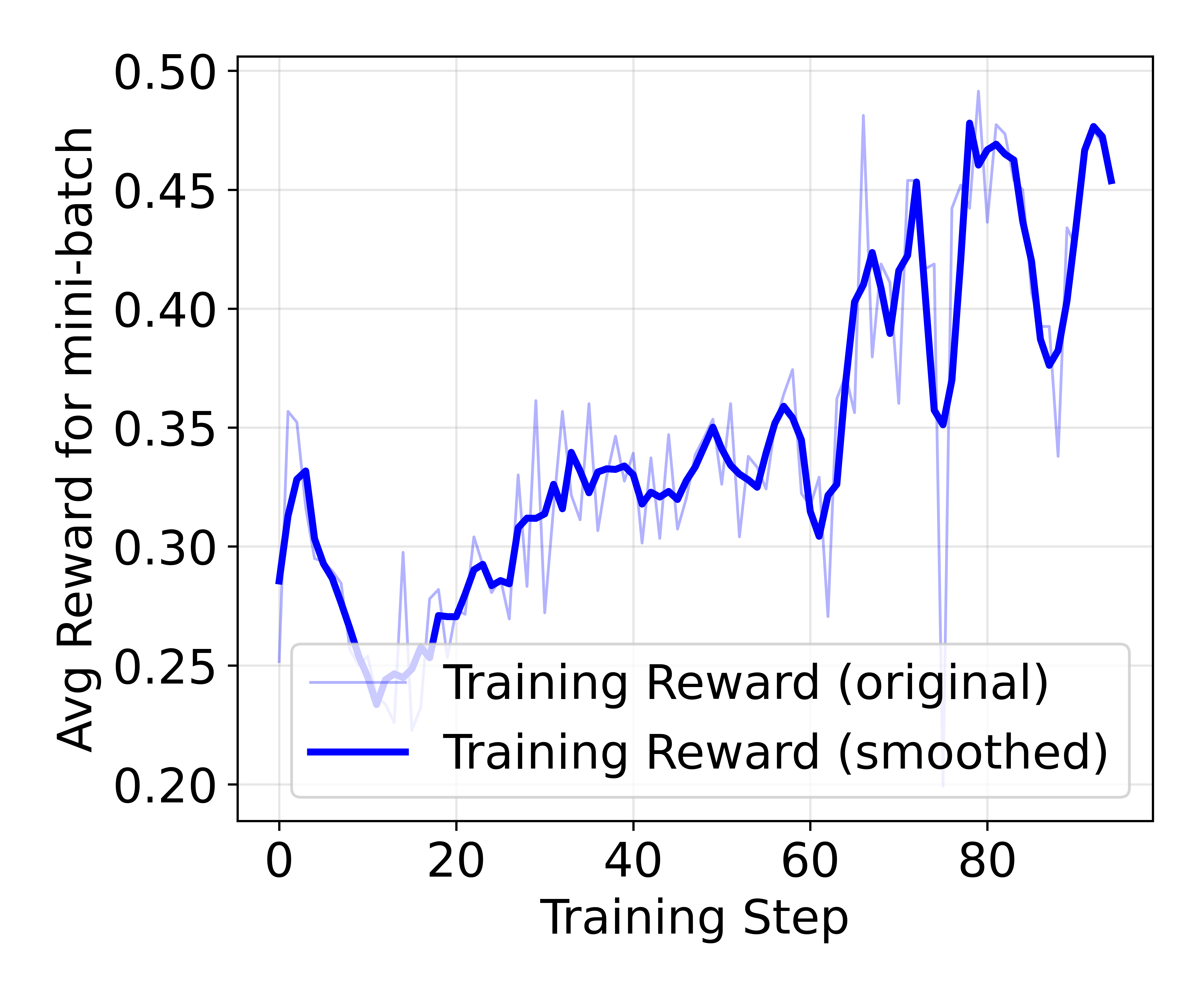}
        \caption{Training Reward}
        % \label{fig:ctm_behav}
    \end{subfigure}
    \hfill
    \begin{subfigure}[b]{0.45\linewidth}
        \includegraphics[width=\linewidth]{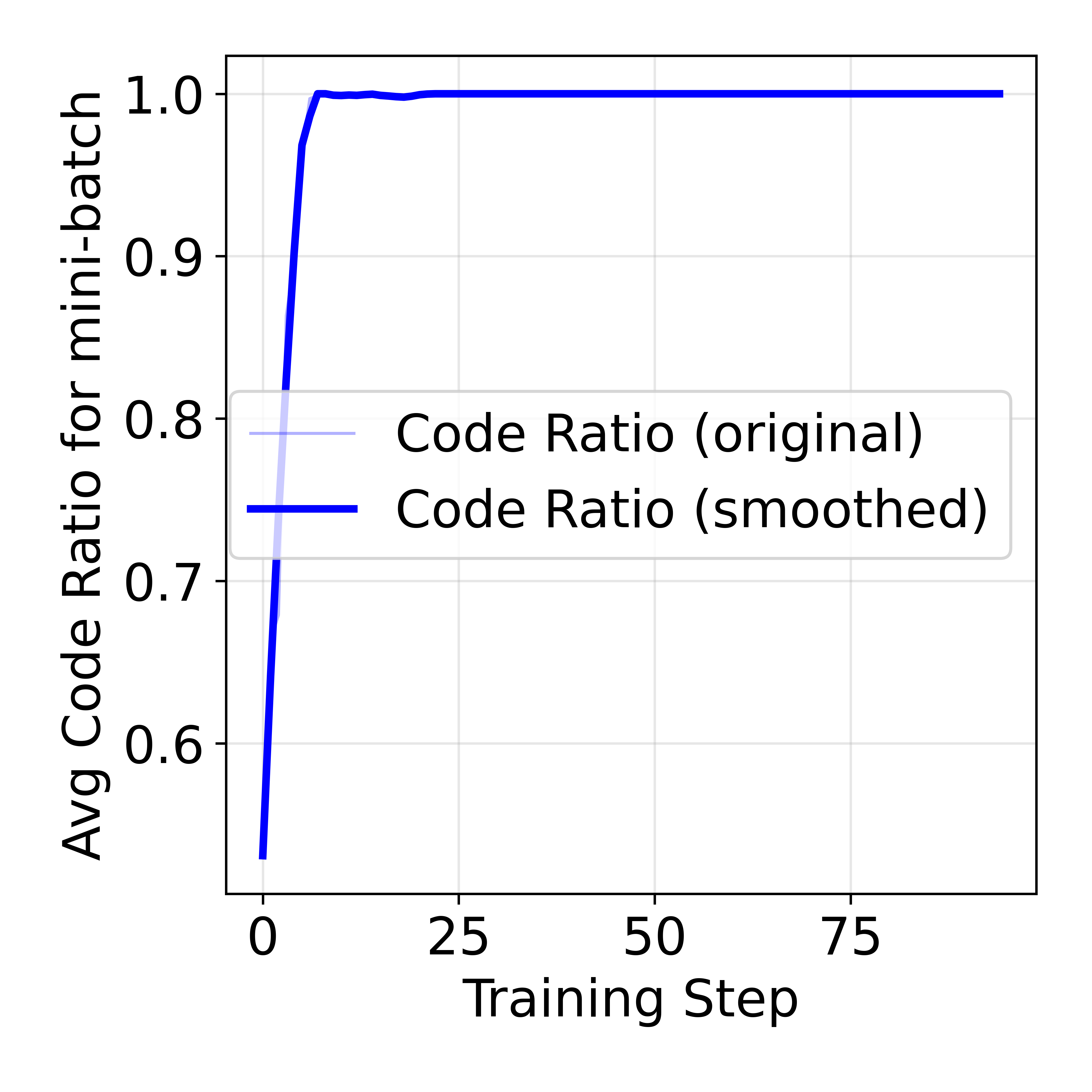}
        \caption{Code Reasoning Ratio}
        % \label{fig:deepseek_behav}
    \end{subfigure}
    \caption{Reinforcement learning training curves for our model.}
    \label{fig:trainingcurve}
\end{figure}

To answer RQ3, we aim to investigate how the added code execution capability enhances the computational thinking model's ability to solve complex problems and affect the reasoning behaviors. 
Following prior work on reasoning behavior analysis \cite{gandhi2025cognitive,zeng2025simplerl}, we leverage \texttt{GPT-4o} to annotate reasoning-related behaviors, including \textit{Backtracking}, \textit{Verification}, \textit{Subgoal Setting}, and \textit{Enumeration}. In addition, we introduce a new annotation criterion to determine whether each behavior is influenced by code execution—resulting in variants such as \textit{Code-Inspired Backtracking}, \textit{Code-Inspired Verification}, etc. For each reasoning output, the annotation model is instructed to assign only one primary behavior category that most accurately characterizes the reasoning process.
The distribution of the behaviors for our computational thinking model is shown in Figure \ref{fig:behavior_comparison} on our sampled 500 questions from \cite{ye2025limo}. 
Case studies for each behavior of our model are shown in Figure \ref{fig:introcase}.

The results reveal that while the overall distribution across the four reasoning behavior categories is relatively similar between our model and other reasoning models \cite{gandhi2025cognitive,zeng2025simplerl}, a key distinction emerges in the degree of \textit{code inspiration}. Within each behavior type, our model demonstrates a significantly higher proportion of code-inspired reasoning, suggesting that the integrated code execution mechanism effectively guides the model toward more accurate and grounded forms of reasoning. 
In contrast, although other natural language based reasoning models such as \texttt{DeepSeek-R1} perform competitively, it lacks a mechanism for executable code interaction. For example, in cases marked as \textit{Verification}, the model can only rely on natural language to evaluate intermediate results—an approach that introduces risks of hallucination and limitations, as previously discussed in Section \S\ref{sec:motivating}.

In addition, we also show the reinforcement learning training curves of our computational thinking model training process, including the training reward curve and the code-reasoning ratio for each mini-batch along the entire training process in Figure \ref{fig:trainingcurve}. 
The reward curve demonstrates a stable upward trend, indicating continuous improvement in the model's overall problem-solving performance during reinforcement learning. 
The code-reasoning ratio—defined as the proportion of rollouts containing \texttt{<code>}...\texttt{</code>} tags—reflects the model's tendency to engage in executable, code-driven reasoning. 
As training progresses, we observe a sharp increase in this ratio, suggesting that the model rapidly learns to leverage code execution as an effective strategy for handling intermediate reasoning steps. 
This provides further evidence that reinforcement learning not only enhances final accuracy but also promotes structured, computationally grounded reasoning behavior.
We further show the case study of our computational thinking model in \S \ref{sec:casestudy} and Figure \ref{fig:casestudy}.

\subsection{RQ4: Ablation Study: Effect of Code Integration}
\label{sec:ablation}
We perform an ablation study to investigate the importance of code integration within our framework. Specifically, we first examine the effect of removing code integration during the training phase, where the model is trained to reason solely using natural language—following the common design of existing language-only reasoning models. In this setting, we retain the same training questions in both the SFT and RL stages as in our original experimental setup.
Additionally, we conduct a separate ablation in which code integration is disabled during the inference stage. In this case, our CTM model is used to produce outputs without invoking the code execution environment, effectively forcing it to rely exclusively on natural language reasoning.
The results of these ablation experiments are summarized in Table \ref{tab:ablation}.
The results consistently indicate that code integration plays a critical role in both the training and inference phases, significantly enhancing the model's reasoning ability.
These results suggest that the interactive code execution environment, combined with computational thinking principles, contributes significantly to the models’ ability to generalize on various complex tasks.

\begin{table}[t]
\centering
\caption{Performance Improvements and Ablation Studies}
\label{tab:ablation}
\label{tab:improvement}
\begin{tabular}{l@{}cc|cc}
\toprule
 & \multicolumn{2}{c}{\textbf{Code}}   & \multicolumn{2}{c}{\textbf{Math (AIME)}} \\
\textbf{Model Variant} & \textbf{LiveCodeBench} & \textbf{CodeContest} & \textbf{2024} & \textbf{2025} \\
\midrule
Qwen2.5-32B-ins & 33.4 & 11.1 & 6.6 & 6.6 \\
\midrule
\textbf{Our CTM} &  & &  &  \\
SFT & 61.6 & 24.2 & 70.0 & 56.6 \\
SFT + RL & \textbf{66.5} & \textbf{28.5} & \textbf{76.6} & \textbf{66.6} \\
\midrule
Training w/o code &  & &  &  \\
SFT & 61.1 & 25.2 & 70.0 & 53.3 \\
SFT + RL  & 63.5 & 26.9 & 76.6 & 60.0 \\
\midrule
Inference w/o code &  & &  &  \\
SFT & 56.3 & 22.1 & 63.3 & 56.6 \\
SFT + RL & 58.5 & 23.8 & 63.3 & 53.3 \\
\bottomrule
\end{tabular}
\end{table}

\section{Discussions}
\subsection{Case Study of Computational Thinking Pattern}
\label{sec:casestudy}

\begin{figure}
    \centering
    \includegraphics[width=\linewidth]{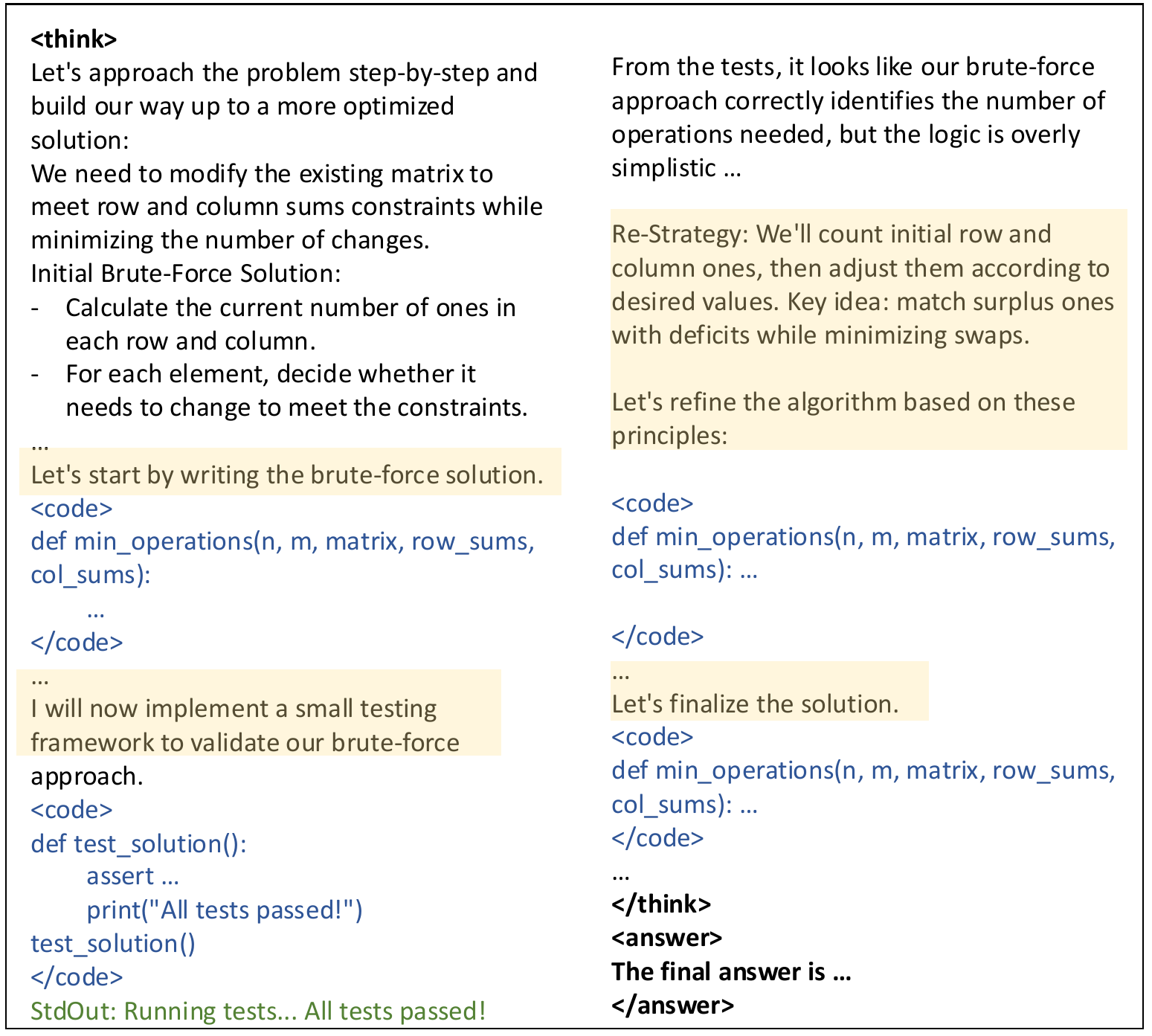}
    \caption{Case Study of the computational thinking pattern for our model. The model shows a \emph{self-imposed validation mechanism} pattern, which enables the model to detect potential errors and enhance solution robustness.}
    \label{fig:casestudy}
\end{figure}
Furthermore, we collect the generated solutions from our Computational Thinking Model (CTM) and manually analyze them to determine whether they exhibit well-known computational thinking patterns, as illustrated in Figure \ref{fig:casestudy}.  
For this challenging coding problem, our CTM demonstrates a \emph{self-imposed validation mechanism} pattern. 
The model first generates a brute-force solution along with corresponding test cases, which are validated against this reference implementation. It then iteratively refines the algorithm into a more optimized version while maintaining correctness through this validation framework. This systematic approach enables the model to detect potential errors and enhance solution robustness.
It aligns with a core principle of computational thinking \cite{wing2006computational}:  
\begin{quote}  
    \textit{Computational thinking is reformulating a seemingly difficult problem into one we know how to solve, perhaps by reduction, embedding, transformation, or simulation.}
\end{quote}

\subsection{Compatibility with Popular RL Algorithms}

The principles underlying our computational thinking model are highly versatile and can be adapted to a wide range of advanced reinforcement learning (RL) techniques. These include offline RL algorithms, such as Reject Sampling \cite{liu2023statistical} and Direct Preference Optimization (DPO) \cite{rafailov2023direct}, as well as online RL algorithms, such as Proximal Policy Optimization (PPO) \cite{schulman2017proximal}, Generalized Policy Optimization (GRPO) \cite{shao2024deepseekmath}, and their advanced variants \cite{yu2025dapo}.
The key to the generality and applicability of our approach lies in its minimally invasive design. Specifically, our method modifies only the rollout behavior during the model's decoding process, without altering the underlying model architecture or the formulation of the training loss function. This decoupling from the foundational components of model training ensures that our computational thinking framework retains compatibility with a variety of RL paradigms and implementations.

\subsection{Threats to Validity}

We summarize several potential threats to the validity of our study and describe the steps taken to mitigate them.

\paragraph{Internal Validity}
Our results may be affected by the model training configuration and the design of the execution environment. The curated reasoning-code traces used for supervised fine-tuning could introduce bias if they overrepresent certain problem types or tool usage patterns. 
We conduct ablation studies to evaluate core components (\eg the code integration for training and inference) in \S \ref{sec:ablation}.

\paragraph{External Validity}
Our evaluation is based on benchmarks such as LiveCodeBench, CodeContests, and AIME. These cover diverse reasoning scenarios, but may not fully capture the complexity of real-world tasks involving long-term planning or multi-modal inputs. Moreover, our computational thinking model currently supports only Python-based code execution, which may limit generalizability to scenarios involving other programming languages or execution modalities.

\paragraph{Construct Validity}
Pass rate and accuracy are the main metrics used to assess model performance. While effective in measuring correctness, it may overlook important qualities such as reasoning efficiency, clarity of intermediate steps, and readability of the reasoning trace. To address this, we provide qualitative analyses of reasoning behaviors to complement quantitative metrics as shown in Figure \ref{fig:introcase} and \ref{fig:casestudy}.

\section{Related Work}

\subsection{Large Reasoning Models}

Recent developments in large language models (LLMs) \cite{gpt4o, liu2024deepseek} have significantly enhanced their ability to perform reasoning tasks. A cornerstone technique driving this progress is Chain-of-Thought (CoT) prompting, first introduced by \cite{wei2022chain}, which decomposes reasoning processes into step-by-step explanations expressed in natural language. CoT prompting has demonstrated remarkable improvements in reasoning tasks, such as mathematical problem-solving, commonsense reasoning, and multi-hop question answering, by enabling models to explicitly articulate intermediate steps.
Building on CoT, research has continued to innovate on reasoning mechanisms. Advanced LLMs like OpenAI-o1 \cite{openaio1} and DeepSeek-R1 \cite{guo2025deepseek} demonstrate the effectiveness of CoT-inspired approaches, achieving state-of-the-art performance on benchmarks requiring systematic decomposition and multi-step reasoning.
Complementing CoT, the Program-of-Thought (PoT) approach \cite{chen2022program, gao2023pal} integrates external computational tools—such as Python interpreters—to simplify and validate reasoning operations. The big difference between our method and the PoT method is that our model can reason in an interactive natural language and code mixed reasoning pattern, which can benefit from both language reasoning models and code execution capability.

\subsection{Tool-Integrated Models}

Building upon the concept of human tool use, recent efforts focus on augmenting LLMs with external computational tools to tackle inherently complex tasks. Tool-integrated reasoning, first introduced to address computationally intensive requirements in domains such as mathematics \cite{chen2022program, schick2023toolformer, yao2023react, zhang2023toolcoder, zhang2023self, li2025start, ma2025automated, paranjape2023art, lu2023chameleon}, provides LLMs access to environments for feedback. 
Early works such as ToolFormer \cite{schick2023toolformer} showed the potential of incorporating tool execution in reasoning processes. 
In addition to traditional tool-integrated frameworks, LLMs have played a key role in developing intelligent agents capable of multi-modal task execution. Representative examples include Auto-GPT \cite{yang2023auto}, CodeAgent \cite{zhang2024codeagent}, and OpenHands \cite{openhands,zhang2025sealign}, which showcase the potential of hybrid systems leveraging reasoning and tool use in complex task-solving workflows. These advances underscore the importance of combining computational capabilities with reasoning strategies, which serve as the foundation of our work.

In recent concurrent work, reinforcement learning (RL) has emerged as a powerful approach for training tool-integrated reasoning models \cite{feng2025retool, qian2025toolrl, li2025torl, openhands}. For instance, the ReTool framework \cite{feng2025retool} applies RL to optimize code interpreter usage specifically for mathematical reasoning tasks, and observes relatively simple tool-use patterns, such as basic code self-correction.

Our research advances this line of concurrent work in two key directions. 
First, we expand the application domain beyond mathematics to encompass more diverse code generation scenarios. 
Second, and more importantly, we introduce a structured methodology in which computational tools are used not only for executing code, but also as instruments to guide the trajectory of reasoning itself. 
Unlike these approaches, the Computational Thinking Model (CTM) embeds computational thinking principles directly into its reasoning architecture during the supervised fine-tuning (SFT) phase and further reinforces this strategy through RL training. 
This design enables greater adaptability, longer-term coherence, and the emergence of higher-level reasoning patterns such as decomposition, recursion, and iterative refinement.
Inspired by the foundational concept of computational thinking \cite{wing2006computational}, CTM reframes problem-solving tasks within a paradigm of structural transformation—through simulation, modularization, and algorithmic abstraction—as illustrated in our case studies in Figures \ref{fig:introcase} and \ref{fig:casestudy}.
These contributions place CTM at the intersection of tool-augmented reasoning and structured computational thinking paradigms. By combining computational precision with cognitive structure, CTM represents a significant step toward building generalizable, interpretable, and robust intelligent systems capable of solving complex, real-world problems.

\section{Conclusion}

In this paper, we propose the computational thinking model (CTM), a novel reasoning model designed to integrate computational thinking principles into large language models. 
By interleaving natural language reasoning with executable code-based reasoning capability, CTM provides a dynamic and interactive reasoning process, making the large reasoning model not just generate text, but think like computer scientists—transforming problems into computational workflows that are robust and verifiable.

\bibliographystyle{plainnat}
\bibliography{sample-base}

\end{document}